\documentclass[apj,twocolumn,twocolappendix,numberedappendix]{openjournal}
\usepackage{newtxtext,newtxmath,enumitem,multirow}
\usepackage[utf8]{inputenc}
\usepackage[T1]{fontenc}
\usepackage[breaklinks,colorlinks,allcolors=blue,citecolor=blue,urlcolor=blue]{hyperref}
\usepackage{orcidlink}
\pdfoutput=1

\def\GRUMPY{\texttt{GRUMPY\,}}

\def\zrei{z_{\rm rei}}
\def\M1500{M_{1500}}
\def\Muv{M_{\rm UV}}
\def\Luv{L_{\rm UV}}
\def\fuv{f_{\rm UV}}
\def\Ls{L_{\star}}
\def\Lt{L_{t}}
\def\phis{\phi_{\star}}

\def\rhouv{\rho_{\rm UV}}
\def\fuv{f_{\rm UV}}

\def\nion{n_{\rm ion}}
\def\dnion{\dot{n}_{\rm ion}}
\def\dNion{\dot{N}_{\rm ion}}
\def\fesc{f_{\rm esc}}
\def\xiion{\xi_{\rm ion}}
\def\aa{\rm \mathring{A}}

\shorttitle{Contribution of dwarf galaxies to reionization}
\shortauthors{Wu \& Kravtsov}

\begin{document}

\title[On the contribution of dwarf galaxies to reionization]{On the contribution of dwarf galaxies to reionization of the Universe\vspace{-1.5cm}}
\author{Zewei Wu\,\orcidlink{0000-0002-7944-2543}$^{1,\star}$}
\author{Andrey Kravtsov\,\orcidlink{0000-0003-4307-634X}$^{1,2,3,\dagger}$\vspace{2mm}}
\affiliation{$^{1}$Department of Astronomy  \& Astrophysics, The University of Chicago, Chicago, IL 60637 USA}
\affiliation{$^{2}$Kavli Institute for Cosmological Physics, The University of Chicago, Chicago, IL 60637 USA}
\affiliation{$^{3}$Enrico Fermi Institute, The University of Chicago, Chicago, IL 60637 USA}
\thanks{$^\star$\href{mailto:wuz25@uchicago.edu}{wuz25@uchicago.edu}, $^\dagger$\href{mailto:kravtsov@uchicago.edu}{kravtsov@uchicago.edu}}

\begin{abstract}
We present estimates of the ultraviolet (UV) and Lyman continuum flux density contributed by galaxies of luminosities from $\Muv\approx -25$ to $\Muv=-4$ at redshifts $5\leq z\leq 10$ using a galaxy formation model that reproduces properties of local dwarf galaxies down to the luminosities of the ultra-faint satellites.
We characterize the UV luminosity function (LF) of galaxies and their abundance as a function of the ionizing photon emission rate predicted by our model and present accurate fitting functions describing them. Although the slope of the LF becomes gradually shallower with decreasing luminosity due to feedback-driven outflows, the UV LF predicted by the model remains quite steep at the luminosities $\Muv\lesssim -14$. After reionization, the UV LF flattens at $\Muv\gtrsim -12$ due to UV heating of intergalactic gas. However, before reionization, the slope of the LF remains steep and approximately constant from $\Muv\approx -14$ to $\Muv=-4$.
We show that for a constant ionizing photon escape fraction the contribution of faint galaxies with $\Muv>-14$ to the UV flux and ionizing photon budget is $\approx 40-60\%$ at $z>7$ and decreases to $\approx 20\%$ at $z=6$. Before reionization, even ultra-faint galaxies of $\Muv>-10$ contribute $\approx 10-25\%$ of ionizing photons. If the escape fraction increases strongly for fainter galaxies, the contribution of $\Muv>-14$ galaxies before reionization increases to $\approx 60-75\%$. Our results imply that dwarf galaxies fainter than $\Muv=-14$, beyond the James Webb Space Telescope limit, contribute significantly to the UV flux density and ionizing photon budget before reionization alleviating requirements on the escape fraction of Lyman continuum photons.
\end{abstract}


\maketitle


\section{Introduction}
\label{sec:intro}
Cosmic reionization of hydrogen was the second major phase transition experienced by the Universe after it became neutral during the epoch of recombination. Modeling details of the reionization process and understanding the main sources of ionizing Lyman continuum (LyC) photons remains an area of active research and debate \citep[see, e.g.,][for reviews]{Robertson.2022,Gnedin.Madau.2022}. 

Two obvious astrophysical sources of LyC radiation are galaxies and active galactic nuclei \citep[AGNs; e.g.,][]{Madau.etal.1999,Faucher_Giguere_etal.2009}. Both observations and cosmological simulations of reionization conclude that young star-forming galaxies contribute the bulk of photons reionizing hydrogen \citep[e.g.,][]{Gnedin.Kaurov.2014, Ma.etal.2015,Robertson.etal.2015,Sharma.etal.2016, Madau.2017, Lewis.etal.2023}, while AGNs play a key role in reionizing helium and maintaining intergalactic medium (IGM) ionized at low redshifts \citep[e.g.,][]{Sokasian.etal.2003}. 

The contribution of galaxies of different luminosities to the hydrogen reionization is still debated. In particular, the contribution of galaxies with UV absolute magnitudes at $\lambda=1500\,\aa$ of $\Muv>-14$ to the UV flux and ionizing photon budget is largely unconstrained by observations and can only be estimated using models. However, theoretical predictions for the UV luminosity function (LF) at these faint magnitudes span a wide range \citep[see, e.g., Figures 12 and 13 in][]{Bouwens.etal.2022}. Some models predict significant flattening of the UV LF or turnover at $\M1500\gtrsim -12$ to $-14$ \citep[e.g.,][]{OShea.etal.2015, Gnedin.2016, Ceverino.etal.2017, Kannan.etal.2022}, while others predict a relatively steep LF down to fainter magnitudes \citep[e.g.,][]{Yue.etal.2016}.

In this study, we aim to predict the evolution of the faint end of the UV LF at $\Muv > -14$ before and during reionization. We employ a new method to sample the evolution of galaxies over the entire relevant range of luminosities, from the brightest observed galaxies down to progenitors of the ultra-faint dwarf galaxies observed around the Milky Way. Galaxy properties are computed using the galaxy formation model of \citet{Kravtsov.Manwadkar.2022}, which was demonstrated to reproduce properties of $L\lesssim L_\star$ galaxies down to ultra-faint dwarf luminosities at $z=0$ \citep{Kravtsov.Manwadkar.2022,Manwadkar.Kravtsov.2022,Kravtsov.Wu.2023}. To account for the stochasticity of star formation rate (SFR) observed in local dwarf galaxies and galaxies at $z>5$, we add a modest level of SFR stochasticity using the method outlined in \citet{Pan.Kravtsov.2023}. We use the star formation and metallicity evolution of model galaxies to compute their AB $\lambda=1500\,\aa$ luminosity and the Lyman continuum photon emission rate using stellar population synthesis, and we predict the UV LF and the LyC photon emission density as a function of galaxy luminosity. We use these functions to estimate the relative contribution of galaxies of $\Muv>-13$ to the UV and ionizing photon flux at $z>5$. 

The paper is organized as follows. 
We describe the galaxy formation model and the method used to estimate properties of the galaxy population across a full range of galaxy luminosities in Section~\ref{sec:modeling}. We present our main results in Section~\ref{sec:results}, compare our results and conclusions to previous studies, and discuss the predicted evolution of the UV and ionizing flux density in Section \ref{sec:discussion}. Our results and conclusions are summarized in Section~\ref{sec:summary}. We provide best-fit values for the predicted number density of ionizing photons produced by galaxies in a given bin of $\Muv$ in the Appendix \ref{app:nion_params}.

Throughout this paper, we assume flat $\Lambda$+Cold Dark Matter ($\Lambda$CDM) cosmology with the mean density of matter in units of the critical density of $\Omega_{\rm m}=0.32$, the mean density of baryons of $\Omega_{\rm b}=0.045$, Hubble constant of $H_0=67.11\,\rm km\,s^{-1}\,Mpc^{-1}$, the amplitude of fluctuations within the tophat spheres of $R=8h^{-1}$ Mpc of $\sigma_8=0.82$, and the primordial slope of the power spectrum of $n_{\rm s}=0.95$. Halo virial masses throughout this study are defined within the radius enclosing density contrast of 200 relative to the critical density at the corresponding redshift.

\section{Modeling high-$z$ galaxy formation}
\label{sec:modeling}

The galaxy formation framework we use in this study is applied to predict galaxy population properties for representative samples of model galaxies at all relevant luminosities down to the UV absolute magnitudes of $M_{1500}\approx -4$ and redshifts $z\in [5,10]$. This is done using samples of halos that follow the expected halo mass function at each considered redshift and halo mass evolution tracks constructed using an accurate approximation for the halo mass accretion rate, as described in \citet{Kravtsov.Belokurov.2024}. 

The key aspect of the galaxy formation model we use in this study at $z\geq 5$ is that it reproduces observed properties of $\lesssim L_\star$ galaxies at $z=0$ down to the faintest ultra-faint dwarf galaxies \citep[][]{Kravtsov.Manwadkar.2022,Manwadkar.Kravtsov.2022,Kravtsov.Wu.2023}. This agreement is not a guarantee that the model would work at high redshifts. Nevertheless, most galaxies at $z>5$ have dwarf halo virial masses ($M\lesssim 10^{11}\,M_\odot$) and $z=0$ give us more confidence that results may be realistic. As we show below, the same model that reproduces the properties of $z=0$ dwarf galaxies also reproduces the observed bright end of the UV luminosity function at $5\leq z \leq 10$.

We briefly outline the main elements of the model relevant to our analysis in Section \ref{sec:grumpy} below. We first describe the modeling of mass assembly histories of halos that host model galaxies, which is the backbone of galaxy formation modeling.

\subsection{Halo evolution model}
\label{sec:hevol}

To model the evolution of halos over the entire range of galaxy luminosities, we first construct large samples of model halos using the following approach \citep{Kravtsov.Belokurov.2024}. We use an accurate cubic spline approximation of the cumulative halo mass function computed using \citet{Tinker.etal.2008} approximation, and use the inverse transform sampling method to generate a random sample of halo masses in a given volume. We use a two-pronged approach to efficiently sample halos hosting galaxies over a very broad range of luminosities. First, we construct halo samples in a series of boxes of different halo masses. When we construct the overall UV or ionizing radiation luminosity function of galaxies, we generate the samples in individual boxes and then stitch the luminosity functions in their ranges of overlap. 

Second, we select a random fraction of halos for modeling, given as a function of halo mass $M_{\rm 200c}$ using: 
\begin{equation}
    f(M_{\rm 200c}) = n \left(\frac{M_{\rm 200c}}{10^9\, M_\odot}\right)^\eta,
\end{equation}
where $n\approx 10^{-5}-5\times 10^{-6}$ and $\eta\approx 1.5-2.35$ provide sufficiently large samples of model galaxies to reliably measure their luminosity functions.
When the luminosity function is computed at a given $z$, each model galaxy is weighted by $f^{-1}(M_{\rm 200c})$. This approach allows us to keep the number of model galaxies reasonably small, while sufficiently sampling galaxies of different luminosities.

Once a halo sample is drawn at a given redshift, $z_{\rm f}$, we use each halo mass as a starting point and construct a halo mass evolution track by integrating the equation of halo mass evolution $\dot{M}_{\rm 200c}=\mu(M_{\rm 200c})$ back in time to $z_{\rm init}=25$
using an accurate approximation for the {\it average\/} halo mass accretion rate $\dot{M}_{\rm 200c}$ of halos of a given mass $M$, derived using analyses of the mass evolution histories of halos formed in cosmological $\Lambda$CDM simulations \citep[see Appendix in][for tests of of the approximation at $z>5$]{Kravtsov.Belokurov.2024}:
\begin{equation}
\dot{M}_{\rm 200c} = 0.3606\, M_\odot\, {\rm Gyr}^{-1}\, M_{\rm 200c}(t)^{1.091}\, t^{-1.8},
\label{eq:dmdt}
\end{equation}
where halo mass $M_{\rm 200c}$ is in $M_\odot$ and time $t$ is in Gyrs. The integrated $M_{\rm 200c}(t)$ is then used to model galaxy evolution from $z_{\rm init}$ to a given $z_{\rm f}$.

\subsection{Galaxy formation model}
\label{sec:grumpy}

The specific implementation of the \GRUMPY galaxy formation model \citep{Kravtsov.Manwadkar.2022} we use is outlined in \citet{Kravtsov.Belokurov.2024}. Briefly, the model solves a system of differential equations that describe the evolution of gas mass, stellar mass, size, and stellar and gas-phase metallicities. It also includes galactic outflows, a model for the gaseous disk and its size, molecular hydrogen mass, star formation, and effects of UV heating during and after reionization on accretion of gas on small-mass halos. 

The \texttt{GRUMPY} model assumes that at all times the ISM follows the exponential radial gas profile $\Sigma_g(R)$, with a half-mass radius proportional to the parent halo virial radius. The model uses molecular gas mass as a proxy for dense cold star-forming gas mass $M_{\rm sf}$ (a fraction of the total gas mass in the galaxy). The molecular fraction $f_{\rm H_2}$ of the total ISM gas mass is estimated using the model of \citet{Gnedin.Draine.2014}. 

The star-forming gas mass, $M_{\rm sf}=f_{\rm H_2}M_{\rm g}$  is converted into stars on a constant depletion time scale $\tau_{\rm sf}$, such that the star formation rate (SFR) is $\dot{M}_\star=(1-R)M_{\rm sf}/\tau_{\rm sf}$, where $R$ is the fraction of gas returned to the interstellar medium in the instantaneous recycling approximation. 
We use the fiducial value of $\tau_{\rm sf}=2$ Gyr typical of nearby galaxies \citep[e.g.,][]{Bigiel.etal.2008}.
The model includes outflows of the ISM gas that are assumed to be proportional to the mean SFR, $\dot{M}_{\rm out}=\eta_w M_{\rm sf}/\tau_{\rm sf}$ with the mass-loading factor $\eta_w$ dependent on the current stellar mass of the galaxy in a way expected in the energy-driven wind models \citep[see][]{Manwadkar.Kravtsov.2022}. 

Note that we assume that scaling of the mass-loading factor $\eta_w$ with stellar mass is the same at different redshifts, as indicated by the results of the FIRE-2 simulations \citep[][]{Muratov.etal.2015,AnglesAlcazar.etal.2017}. Note, however, that the star formation is not assumed to be the same at all redshifts, as the amount of star-forming gas in the model depends on the gas metallicity and UV field and both are different at high z and this is taken into account. We do assume that depletion time of gas stays constant, while it may be redshift dependent. We find, however, that effects of changing depletion time with redshift within reasonably limits is quite small at z<10, as was also found in cosmological simulations \citep[e.g.,][]{Schaye.etal.2010}.

We do not take into account the effects of mergers on the stellar populations of model galaxies. 
In the dwarf galaxy regime mergers have a negligible effect on the stellar masses of galaxies \citep{fitts_etal18} due to the steep $M_{\rm halo}-M_\star$ relation. Although major mergers can change the stellar mass and size of dwarf galaxies \citep{Rey.etal.2019,tarumi_etal21}, such mergers are quite rare and do not affect average scaling relations. Note also that the model for the galaxy half-mass radius provides a good match to the distribution of half-mass radii of observed dwarf galaxies at $z=0$. 

The gas accretion onto galaxies is modulated by the mass and redshift dependent factor accounting for the UV heating effect of intergalactic gas, as described in \citet{Kravtsov.Manwadkar.2022}. To illustrate the effects of this heating on the UV and ionization luminosity functions of dwarf galaxies in our analysis, we will consider models with reionization redshifts $\zrei = 6$ and $\zrei = 8.5$. The former is close to the redshift at which our Universe was reionized \citep[see][for reviews]{Gnedin.Madau.2022,Robertson.2022} and is our fiducial value. The $\zrei=8.5$ value is used to illustrate how the evolution of the luminosity functions changes if a given region of the Universe is reionized earlier. 

For each halo track produced as described above, the galaxy formation model is integrated from $z_{\rm init}=25$ to the final redshift $z_{\rm f}=5, 6, 7, 8, 9, 10$, producing evolution of stellar mass, star formation rate, etc. The basic model described above, however, uses mean mass assembly history for a halo of a given mass at $z_{\rm f}$, while real halos of a given will exhibit a scatter in their assembly histories. In addition, the model assumes the same star formation depletion time for all galaxies, while observational estimates of $\tau_{\rm sf}$ vary significantly from galaxy to galaxy. In addition, the model does not include modeling of the processes that can result in a significant SFR stochasticity, such as the formation and destruction of individual star-forming regions \citep[e.g.,][]{Tacchella.etal.2020,Iyer.etal.2020,Sugimura.etal.2024}. 

To account for these different sources of scatter in SFR in a controlled manner, we add stochasticity to the mean SFR computed by the model using the method described in \citet{Pan.Kravtsov.2023}. 
Namely, the mean SFR in the model at a time $t_n$ is perturbed as $\dot{M}_{\star,\rm stoch} = \dot{M}_{\star}\times 10^\Delta$,
where $\Delta$ is a correlated random number drawn from the Gaussian pdf with zero mean and unit variance, and multiplied by $\sqrt{P(k)}=\sqrt{{\rm PSD}(f_k)/T}$, where wavenumber $k$ corresponding to frequency $f_k$ is defined as $k=f_k T$ and where $T$ is the duration of galaxy evolution track.

We follow  \citet{Caplar.Tacchella.2019} and use the PSD of the form ${\rm PSD}(f) = \sigma^2_\Delta[1+(\tau_{\rm break}f)^\alpha]^{-1}$,
where $\sigma_\Delta$ characterizes the amplitude of the SFR variability over long time scales and  $\tau_{\rm break}$ characterizes the timescale over which the random numbers are effectively uncorrelated. Parameter $\alpha$ controls the slope of the PSD at high frequencies (short time scales). 
In our models, we fix the slope $\alpha$ and $\tau_{\rm break}$ to the values $\alpha=2$ and $\tau_{\rm break}=100$ Myr, which are physically motivated by the time scales of gas evolution and star formation in giant molecular clouds in a typical ISM \citep[see][for a detailed discussion]{Tacchella.etal.2020}, as well as $\sigma_\Delta=0.1$ consistent with the typical amount of SFR stochasticity in host halos of observed galaxies at $z\approx 5-10$. The corresponding scatter in the UV absolute magnitude is $\sigma_{\rm M_{\rm UV}}\approx 0.75$ \citep[see][for further exploration of the effects of stochasticity in the context of the model we use]{Kravtsov.Belokurov.2024}, which is broadly consistent with the UV absolute magnitude fluctuations estimated in high-resolution zoom-in cosmological simulations at the same redshifts \citep{Pallottini.Ferrara.2023,Sun.etal.2023}.

\subsection{Computing UV and ionizing radiation luminosities}
\label{sec:comp_uv_ion}

The monochromatic luminosity of model galaxies at $\lambda=1500\,{\aa}$ is computed using a tabulated grid of luminosities, $L_{1500}$, for stellar populations of a given age and metallicity using the Flexible Stellar Population Synthesis model v3.0 \citep[FSPS,][]{Conroy.etal.2009,Conroy.Gunn.2010} and its Python bindings, \textsc{Python-FSPS}. The table is then used to construct an accurate bivariate spline approximation to compute $L_{1500}$ for stellar populations of a given age and metallicity. 
We use the table and finely spaced time outputs of the model to compute the integral $L_{1500}$ due to all stars formed by the current time taking into account the evolution of stellar mass and stellar metallicity. 

The emission rate of the Lyman continuum $\lambda<912\,\aa$ photons, $\dNion$, is sensitive to the effects of binary stars (while flux at $\lambda=1500\,\aa$ is not), which are not included in the FSPS. We thus computed $\dNion$ self-consistently using the tables of ionizing flux for single-age stellar populations from the BPASS version 2.3 package \citep{Byrne.etal.2022}, which take into account effect of binary stars, and the evolution of stellar mass and metallicity computed by the galaxy formation model.

To account for dust effects, which are expected to affect the brightest galaxies in the UV at $z\lesssim 7$, we use a second-order polynomial approximation to the simulation results of \citet[][see their Fig. 7]{Lewis.etal.2023}: $A_{1500} = -0.07M_{1500,Z} + 0.05 M_{1500,Z}^2$, where $M_{1500,Z}= 2 + 10.53\log_{10}(0.043/Z_{\rm gas})$ and $Z_{\rm gas}$ is metallicity of gas in solar units, which we assume to be $Z_\odot=0.015$. 

\begin{figure}
   \centering {
   \includegraphics[width=0.49\textwidth]{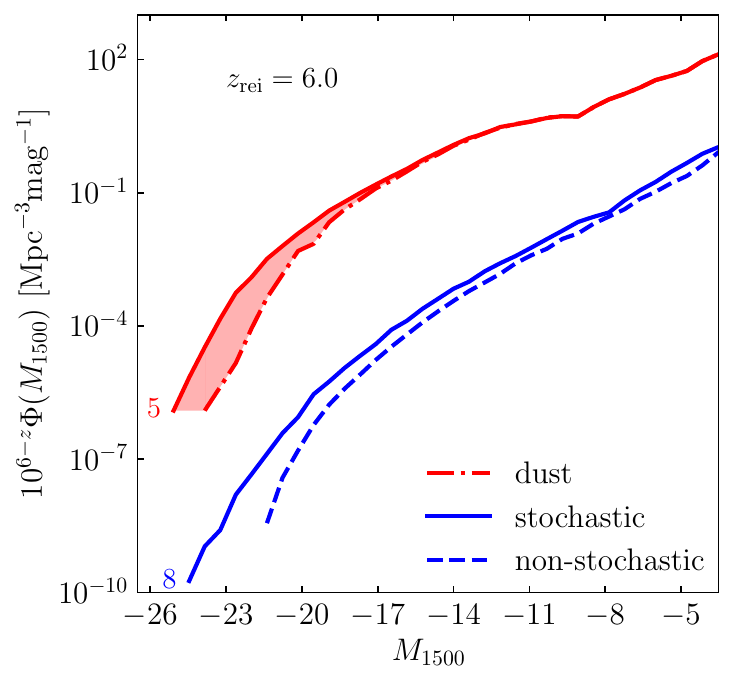}
   }
   \caption{
   UV LFs at redshifts $z=5$ and $8$ over the entire range of galaxy luminosities from $\Muv\approx -25$ to $\Muv=-4$, assuming that reionization ends at $\zrei=6$. At each redshift, the two curves illustrate the effects of dust ($z=5$; solid line is our model without accounting for dust, while dot-dashed line is the LF after applying dust extinction) and the effect of adding a small level of SFR stochasticity to the model results ($z=8$; the dashed curve shows the model without SFR stochasticity; solid line is the LF in the model with stochasticity). The LFs are shifted by a factor $10^{6-z}$ for clarity.
   }
   \label{fig:dust_stoch}
\end{figure}

The luminosities $L_{1500}$ and $\dNion$ for each model galaxy are computed at the final redshifts $z_{\rm f}=5,6,7,8,9,10$. To construct the corresponding luminosity functions at each $z_{\rm f}$, we construct a weighted histogram of luminosities using weights $f^{-1}(M_{\rm 200c})L^{-3}_{\rm box}$. As noted above, we first construct luminosity functions in individual volumes of a series of increasing $L_{\rm box}$ and then stitch the luminosity functions in the regions of overlap, so that the combined LF spans the full range of galaxy luminosities down to $M_{1500}= -4$. 

Our reasons for choosing the lower luminosity limit of $\Muv=-4$ are twofold. First, in our model, this luminosity corresponds to a stellar mass of $\approx 300\, M_\odot$. For this and smaller masses, stochastic sampling effects of the initial mass function are expected to be significant \citep[e.g.,][]{daSilva.etal.2012}, which the feedback prescription in our model does not account for. Feedback should be smaller when fewer massive stars per unit stellar mass are formed, and the lack of such stars will also suppress UV luminosity. Second, the halo mass corresponding to $\Muv=-4$ at $z<10$ is $M_{\rm 200c}\approx 5\times 10^6\, M_\odot$, which is close to the minimum halo mass of $\approx 2-5\times 10^6, M_\odot$ that can accrete gas at $z<15$. Gas accretion is expected to be suppressed in halos of smaller mass due to the bulk motion of baryons relative to dark matter and due to expected radiative heating \citep[see, e.g., Fig. 9 in][]{Nebrin.etal.2023}. 

The exact mass threshold for galaxy formation depends on the specific value of the relative streaming velocity of baryons relative to dark matter and the evolution of the Lyman-Werner UV background. The galaxy formation model we use does not account for these effects and thus cannot reliably model properties of $\Muv>-4$ galaxies. However, even if we use the slope of UV LF estimated at $\Muv=-4$ and extrapolate it to $\Muv=-1$, which corresponds to $M_\star\approx 10-50\, M_\odot$ and should have greatly suppressed $L_{1500}/M_\star$ due to deficiency of massive stars, we estimate that the contribution of galaxies with $\Muv>-4$ should be no larger than $\approx 10\%$.   

The UV luminosity functions produced with this method over a range of absolute magnitudes $-25\lesssim M_{1500}\lesssim -4$ at redshifts $z=5$ and $z=8$ is illustrated in Figure~\ref{fig:dust_stoch}. For $z=5$ we show two LF curves computed with and without accounting for the effects of dust discussed above. Dust primaril

y reduces luminosities of bright galaxies of $M_{1500}\lesssim -19$ and its effects become small at all luminosities at $z>7$.  
The two curves at $z=8$ show model UV LFs for the base model without SFR stochasticity and the model in which a small amount of SFR stochasticity with $\sigma_\Delta=0.1$ \citep[required to match observed UV LF at these redshifts; see][]{Kravtsov.Belokurov.2024} is added. This level of SFR stochasticity modifies the shape of the bright end of UV LF at $\Muv\lesssim -17$. 

As the figure shows, the effects of dust and adding SFR stochasticity partly offset each other. Based on the uncertainty and the relatively small influence the dust effects have on our conclusions, we do not include them in our calculations of the relative contribution of galaxies of different luminosities to the total UV and ionizing flux of galaxies. Given that the effect of dust is larger for brighter galaxies, this implies that we underestimate the relative contribution of dwarf galaxies to the UV flux, making our estimates of their contribution a conservative lower limit. However, below we provide fits to the UV LFs and ionizing flux functions in our model and these can be used to recompute the fractional contribution of dwarf galaxies to the UV and Lyman-continuum photon budgets for a specific model of dust attenuation.

\section{Results}
\label{sec:results}

\subsection{UV luminosity functions}
\label{sec:uv_luminosity_function}

\begin{figure}
   \centering {
   \includegraphics[width=0.49\textwidth]{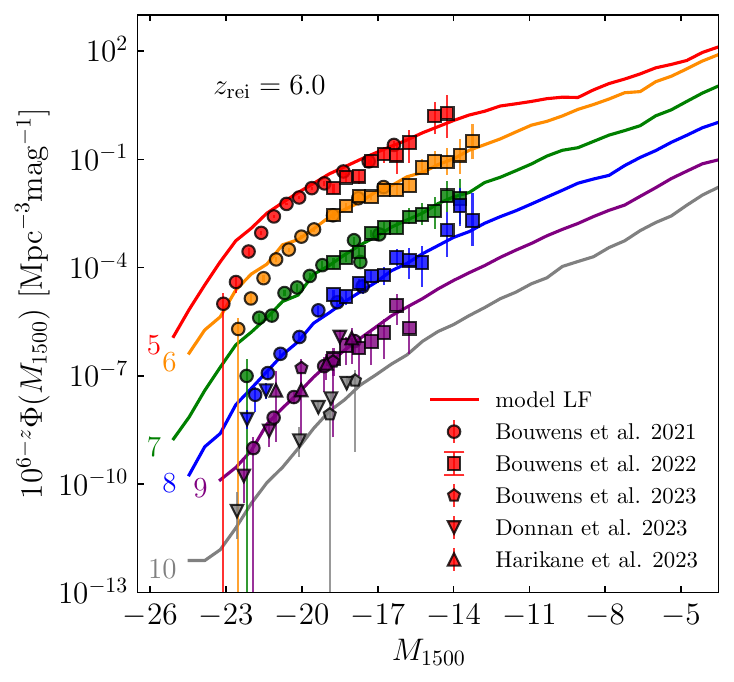}
   }
   \caption{
   Rest-frame UV luminosity function of galaxies at $z\in[5, 10]$ in the model with the end of reionization at $\zrei = 6$ and a small amount of SFR stochasticity of $\sigma_\Delta=0.1$ (see Section \ref{sec:grumpy}). Effects of dust are not included in the model LFs. The LFs at different redshifts are displaced vertically by a factor of $10^{6-z}$ for clarity. The different symbols show observational estimates of the UV LF in recent studies that used HST and JWST observations \citep{Bouwens.etal.2021, Bouwens.etal.2022, Bouwens.etal.2023, Donnan.etal.2023, Harikane.etal.2023, Harikane.etal.2024, Perez.Gonzalez.etal.2024}. Note that before reionization (i.e. $z \gtrsim 6$), the slope of the LF even at the faintest magnitudes remains as steep as that of $M_{1500}\approx -14$. {\it Bottom panel}: $\log_{10}$ of model and observation, shown on the same magnitude range.
   }
   \label{fig:UVLF_obsv}
\end{figure}

Figure~\ref{fig:UVLF_obsv} shows the UV ($\lambda=1500\ \aa)$ luminosity function of galaxies at $z=5, 6, 7, 8, 9, 10$ in the model with the end of reionization at $\zrei = 6$ and a small amount of SFR stochasticity of $\sigma_\Delta=0.1$ (see Section \ref{sec:grumpy}). It also shows observational estimates of the UV LF in recent studies that used HST and JWST observations. The model matches these quite well at the range of luminosities probed by observations. Differences at $z\leq 7$ and $M_{1500}\lesssim -20$ are likely due to dust effects that have a similar magnitude at these luminosities and redshifts (see Fig. \ref{fig:dust_stoch}).  

\begin{figure}
   \centering {
   \includegraphics[width=0.49\textwidth]{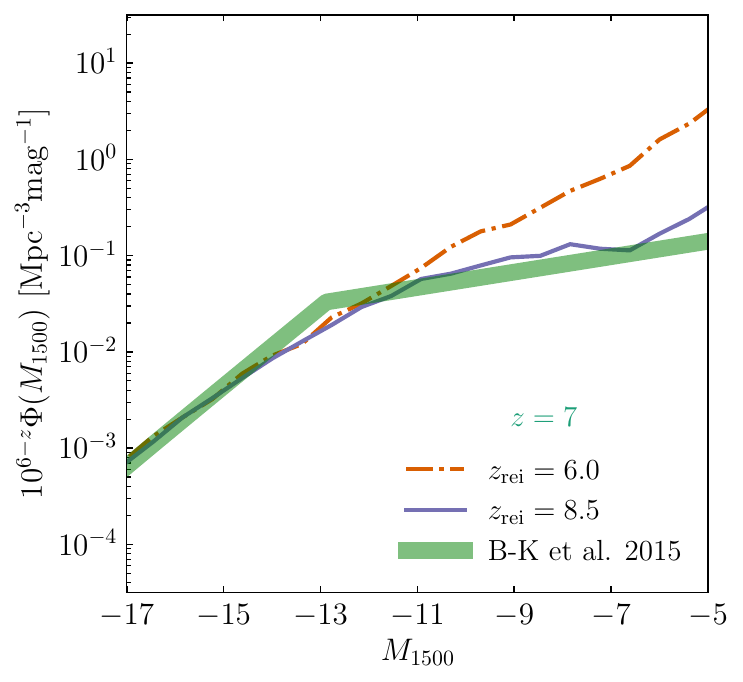}
   }
   \caption{
   Comparison of rest-frame $z=7$ UV LFs in models with reionization redshifts $\zrei = 6$ and $8.5$. The green curve shows the LF function estimated at $z\approx 7$ based on the star formation histories of the Local Group dwarf galaxies  \citep{BoylanKolchin.etal.2015}. The Local Group reconstructed LF is in agreement with the model with $\zrei=8.5$, which indicates that the Local volume was reionized around this redshift.
   }
   \label{fig:UVLF_BK_turnover}
\end{figure}

Agreement with UV LF estimates at $z\in[5, 10]$ at $M_{1500}<-14$ and the fact that the model reproduces properties of $z=0$ dwarf galaxies well \citep{Kravtsov.Manwadkar.2022}, including the luminosity function of Milky Way satellites down to the ultra-faint magnitudes \citep{Manwadkar.Kravtsov.2022}, means we can plausibly expect that the model UV LF can faithfully describe the evolution of the luminosity function at fainter magnitudes.

This statement is supported by the comparison with the reconstruction of the UV LF of the progenitors of the Local Group dwarf galaxies at $z\approx 7$ of \citet{BoylanKolchin.etal.2015} shown in Figure \ref{fig:UVLF_BK_turnover}. This LF reconstruction was done using observational estimates of the star formation histories of dwarf galaxies measured from their color-magnitude diagrams \citep{Weisz.etal.2014,Weisz.etal.2019,Weisz.BoylanKolchin.2017} and their corresponding $M_{1500}$ at $z\approx 7$. The function is a combination of the Schechter form with $M_\star = -21.03$, $\phi_\star = 1.57 \times 10^{-4}\,\rm mag^{-1}\, Mpc^{-3}$, $\alpha = -2.03$ at $M_{1500}<-13$ \citep{Finkelstein.etal.2015} and a power law $\phi\propto L_{1500}^{-1.2}$ at $M_{1500}\geq -13$. As shown by \citet{BoylanKolchin.etal.2015}, the flattening of the slope at $M_{1500}$ is required by the observed abundance of Local Group dwarf galaxies with such faint estimated $z=7$ absolute magnitudes. 

Figure~\ref{fig:UVLF_BK_turnover} shows the UV LF of our model galaxies for the models with the end of reionization at $\zrei=6$ and $\zrei=8.5$. The $\zrei=8.5$ model is in good agreement with the faint-end UV LF deduced by \citet{BoylanKolchin.etal.2015}, because by $z=7$ galaxies of $M_{1500}\gtrsim -13$ become affected by the UV heating, while in the $\zrei=6$ this occurs only at $z\lesssim 6$. This also agrees with the analysis of \citet{Manwadkar.Kravtsov.2022}, which used the same model as in our analysis to show that the $z=0$ luminosity function of the Milky Way satellites favors reionization at $\zrei\approx 8-9$. 

Note that the Lagrangian volume which collapsed into the volume containing nearby dwarf galaxies is generally expected to reionize at $z\gtrsim 7$ \citep{Zhu.etal.2019,Ocvirk.etal.2020,Trac.etal.2022} -- earlier than the overall reionization of the Universe, which occurred at $\zrei\approx 6$ \citep{Gnedin.Madau.2022,Robertson.2022}. Thus, the flattening of the UV LF at $M_{1500}\gtrsim -13$ exhibited by the local dwarf galaxies does not imply that the mean UV LF of $z=7$ galaxies in the Universe should have a similar flattening. 

Before reionization, UV LFs shown in Figure \ref{fig:UVLF_obsv} become gradually shallower with decreasing luminosity (increasing $\Muv$) down to $\Muv\approx -14$, while at fainter magnitudes the slope stays approximately constant down to $\Muv=-4$.  The value of the slope $\alpha$ in $dn/dL\propto L^{\alpha}$ at these faint luminosities is $\alpha\approx -1.7$, which is quite steep.

The flattening at brighter magnitudes and the constant slope at fainter ones is a result of the specific feedback-driven outflow prescription adopted in the model which was tested and calibrated using the mass-metallicity relation of local dwarf galaxies and luminosity function of the Milky Way satellites. 
In what follows, we will present analytical fits to the luminosity functions predicted in our model both for the UV luminosity at $\lambda=1500\,\aa$ and for the emission rate of ionizing photons. 

\subsection{Modified Schechter function fit}
\label{sec:schechter_fit}

\begin{figure}
   \centering {
   \includegraphics[width=0.49\textwidth]{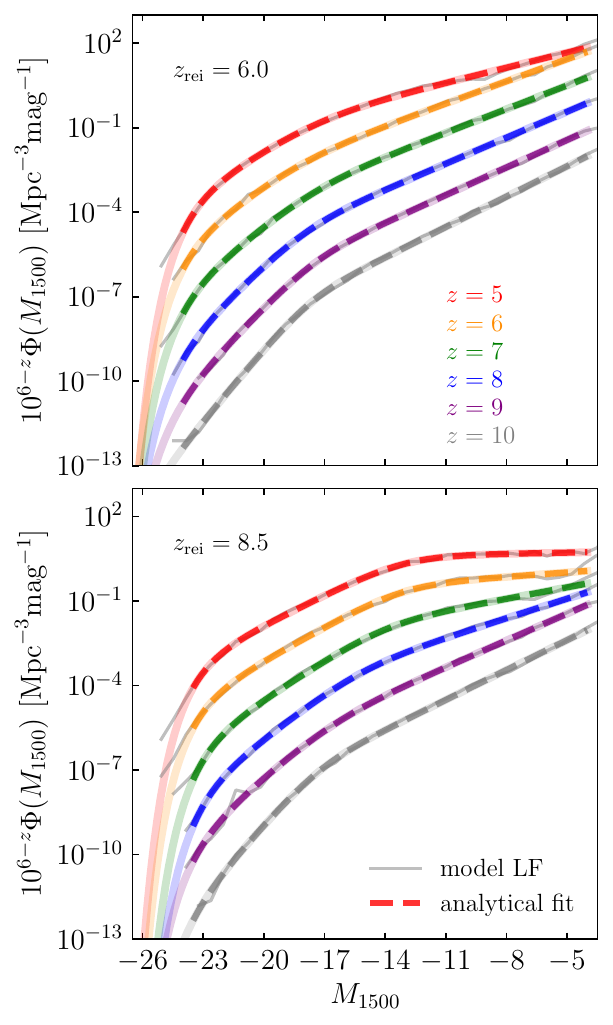}
   }
   \caption{
   Least-squares fit on the model LFs at $z=5, 6, 7, 8, 9, 10$ with \citet{Jaacks.etal.2013}'s modified Schechter function shown in Equation~\ref{eq:jaacks}, fitted on $-26 \le \M1500 \le -4$ (see the best-fit parameters in Table~\ref{tab:jaacks_params}). The two panels show fits to the LFs in models with $\zrei = 6.0$ and $\zrei = 8.5$, respectively. The faint-end behavior, including the steep slope before reionization and the flattening in post-reionization redshifts $z < \zrei$, are accurately reflected by the modified Schechter fit.
   }
   \label{fig:schechter_fit}
\end{figure}

\begin{table}
    \centering
	\caption{
 Best-fit parameters for \citet{Jaacks.etal.2013}'s modified Schechter function to our stochastic UV LF at $z\in[5,10]$. The last column $\alpha - \beta$ shows the effective faint end LF slope.
 }
	\begin{tabular}{ccccccc}
		\hline\hline\\[-2mm]
		$z$ & $\log_{10}\phis$ & $M_{1500, \star}$ & $M_{1500, t}$ & $\alpha$ & $\beta$ & $\alpha-\beta$ \\
            & ${\rm Mpc^{-3}}$ & & & & & \\[1mm]
		\hline\\[-2mm]
        \multicolumn{7}{c}{$\zrei = 6.0$} \\[1mm]
            5 & -2.531 & -22.94 & -17.18 & -0.445 & 0.675 & -1.121 \\
            6 & -3.334 & -23.20 & -18.20 & -0.655 & 0.672 & -1.327 \\
            7 & -3.616 & -23.30 & -17.50 & -0.699 & 0.754 & -1.452 \\
            8 & -4.144 & -23.58 & -17.84 & -0.769 & 0.947 & -1.715 \\
            9 & -4.589 & -23.90 & -17.48 & -0.814 & 1.026 & -1.840 \\
            10 & -5.665 & -25.32 & -17.72 & -0.898 & 1.214 & -2.111 \\[1mm]
        \multicolumn{7}{c}{$\zrei = 8.5$} \\[1mm]
            5 & -0.501 & -22.61 & -13.04 & -0.031 & 0.858 & -0.889 \\
            6 & -0.884 & -22.59 & -13.19 & -0.128 & 0.869 & -0.997 \\
            7 & -1.850 & -22.36 & -14.06 & -0.336 & 0.824 & -1.160 \\
            8 & -2.913 & -22.39 & -15.21 & -0.574 & 0.787 & -1.361 \\
            9 & -4.072 & -22.78 & -16.59 & -0.790 & 0.762 & -1.552 \\
            10 & -4.777 & -23.07 & -17.19 & -0.884 & 1.097 & -1.981 \\[1mm]
    \hline
    \label{tab:jaacks_params}
    \end{tabular}
\end{table}

We approximate model UV luminosity functions with the modified Schechter functional form of \citet{Jaacks.etal.2013}:
\begin{equation}
    \Phi(L) = \phis {\left(\frac{L}{\Ls}\right)}^{\alpha} \exp{\left(-\frac{L}{\Ls}\right)} {\left[1+{\left(\frac{L}{\Lt}\right)}^{\beta}\right]}^{-1},
    \label{eq:jaacks}
\end{equation}
where $\phis$ and $\Ls$ are the normalization and characteristic luminosity of the bright end, respectively. 
Compared to the Schechter form, which has a fixed faint-end slope $\alpha$, this form has a slope that can become progressively shallower or steeper around $\Lt$ and reaches the asymptotic slope of $\alpha-\beta$ at $L\ll \Lt$.  

We determine the best-fit parameters of the function by minimizing the least-squares differences between the functional form and model UV LF converted from the luminosity to absolute magnitude $\Muv$ using the conversion
\begin{eqnarray}
\Muv &=& -2.5\log_{10} \frac{L_{\rm UV}}{4\pi (10\,{\rm pc})^2} - 48.6\nonumber\\ &=& -2.5\log_{10} L_{\rm UV} + 51.59,
\end{eqnarray}
where $L_{\rm UV }$ is the luminosity density at $\lambda=1500\,\aa$ in $\rm egs\,s^{-1}\, Hz^{-1}$. 

The best-fit parameters for different redshifts in models with $\zrei=6$ and $\zrei=8.5$ are presented in Table~\ref{tab:jaacks_params} and the fits are compared to the computed model UV LFs in Figure~\ref{fig:schechter_fit}. The figure shows that in both $\zrei$ models the functional form provides an excellent description of the model luminosity functions at all $z$ and over the entire range of luminosities. 

Figure~\ref{fig:schechter_fit} also illustrates the effect of $\zrei$: the $\zrei=8.5$ model shows significant flattening at the faint end for redshifts $z \lesssim 8.5$ due to suppression of accretion caused by the UV heating of the intergalactic medium during and after reionization. The flattening is also reflected in the lower $\alpha-\beta$ values at these redshifts in Table~\ref{tab:jaacks_params}. In the $\zrei=6$ model, such flattening also occurs after reionization and is apparent only in the $z=5$ LF.  

\subsection{Fraction of UV emission from galaxies of different $M_{1500}$}
\label{sec:frac_uv}

We use the functional LF fits of Equation \ref{eq:jaacks} to calculate the fraction of UV luminosity per unit volume emitted by galaxies brighter than a given $\Muv\leq -4$:

\begin{equation}
   \fuv (<\Muv) = f_{\rm norm}\,\int_{-\infty}^{\Muv} \Luv\,\frac{dn}{d\Muv}\, d\Muv,
   \label{eq:f_uv}
\end{equation}
where normalization is 
\begin{equation}
   f_{\rm norm}=\left[\int^{-4}_{-\infty} \Luv\,\frac{dn}{d\Muv}\, d\Muv\right]^{-1}.
   \label{eq:fnorm}
\end{equation}

\begin{figure}
   \centering {
   \includegraphics[width=0.49\textwidth]{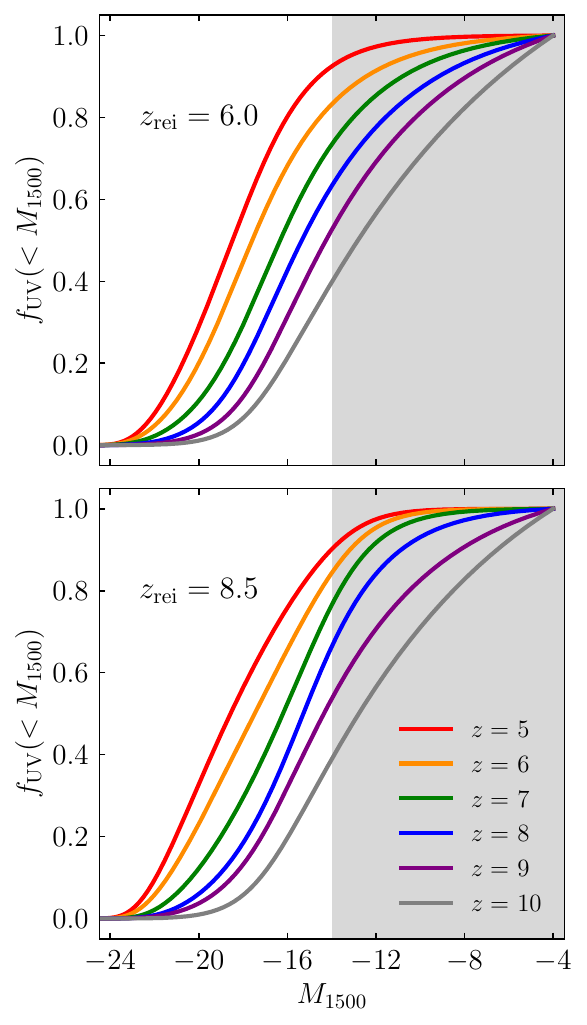}
   }
   \caption{
   The fraction of UV flux contributed by galaxies brighter than a given $\Muv$ for two reionization models ($\zrei = 6$ and $8.5$ respectively) across redshifts $z\in[5,10]$. The fraction is computed by integrating the UV LF over the luminosity range and dividing by the integral over the entire luminosity range (See Equation~\ref{eq:f_uv} and~\ref{eq:fnorm}). Both plots show results for the model with SFR stochasticity. The figure shows that dwarf galaxies contribute a significant fraction of the UV flux ($\approx 55-65\%$ at $\Muv>-14$), especially at higher $z$.
   }
   \label{fig:fractional_flux}
\end{figure}
Figure~\ref{fig:fractional_flux} shows $f(<\Muv)$ for both $\zrei=6$ and $\zrei=8.5$ models for $z=5, 6, 7, 8, 9, 10$.
For both choices of $\zrei$, we see a significant contribution of UV flux from dwarf galaxies with $M_{1500} > -14$. The fraction of the UV flux they contribute is  $\approx 60\%$ at $z=10$ and declines with decreasing $z$ to $\approx 10\%$ at $z=5$. Such decrease is due to 1) changing shape of the bright end of UV LF which becomes shallower with decreasing $z$ due to the continuing buildup of massive halos and galaxies and 2) flattening of the UV LF at $M_{1500}\gtrsim -13$ due to the UV heating after reionization (see discussion in Section \ref{sec:schechter_fit}). We can see the effect of the latter in the rapid steepening of $f_{\rm UV}$ at $z=8$ compared to $z=9$ for the model with $\zrei=8.5$ shown in the bottom panel of Figure \ref{fig:fractional_flux}. 
The UV flux contribution of $M_{1500} > -14$ galaxies in the model with $\zrei=8.5$ right before reionization is $\gtrsim 50-60\%$, while in the $\zrei=6$ model it is $\gtrsim 20-30\%$. The contribution of dwarf galaxies to the global UV flux is thus higher at higher $z$, but is still substantial even at $z\approx 6-7$ in the $\zrei=6$ case. 

Note that due to the reionization-induced flattening of UV LF at $z\gtrsim\zrei$ for $\Muv\lesssim -13$, there is no divergence of the integral of the LF as it is integrated to lower luminosities, as occurs for the Schechter form that approximates UV LF of bright galaxies at $\Muv<-13$. In our model LF a sharp flattening or cutoff occurs only at $\Muv>-4$. Before reionization, $f_{\rm UV}$ continues to increase with decreasing luminosity down to $\Muv=-4$ due to the constant and relatively steep slope of the predicted faint end function. However, at $\Muv>-4$ galaxies in our model have stellar masses $M_\star\lesssim 200\,M_\odot$ and form in halos of $\lesssim 5\times 10^6\,M_\odot$. Halos with virial mass of $M_{\rm 200c}<2\times 10^6\, M_\odot$ do not form stars, as expected in models of gas cooling \citep{Nebrin.etal.2023}, which provides a natural LF cutoff and prevents divergence of $f_{\rm UV}$ at $\Muv<-4$. Note also that \citet{Nebrin.etal.2023} show that at $z<8$ the galaxies in halos with masses $M_{\rm 200c}<10^8\,M_\odot$, which corresponds to $\Muv\gtrsim -8$ in our model, cannot accrete gas due to radiative heating. This does not necessarily mean that UV LF flattens strongly at these faint luminosities immediately at $z<8$ as the galaxies can continue to form stars using the gas accreted at earlier epochs. The effect of such gas suppression will be felt at $z<7$ when effects of UV heating start to affect UV LF anyway.  

\subsection{Ionizing emission fraction from galaxies of different $M_{1500}$}
\label{sec:frac_ion}

\begin{figure*}
   \centering {
   \includegraphics[width=0.49\textwidth]{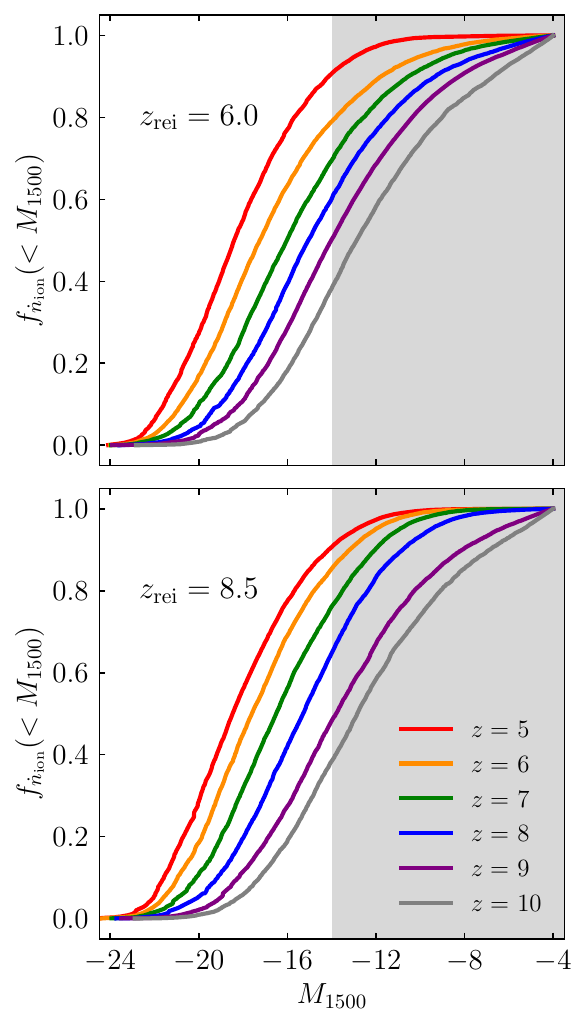}
   \includegraphics[width=0.49\textwidth]{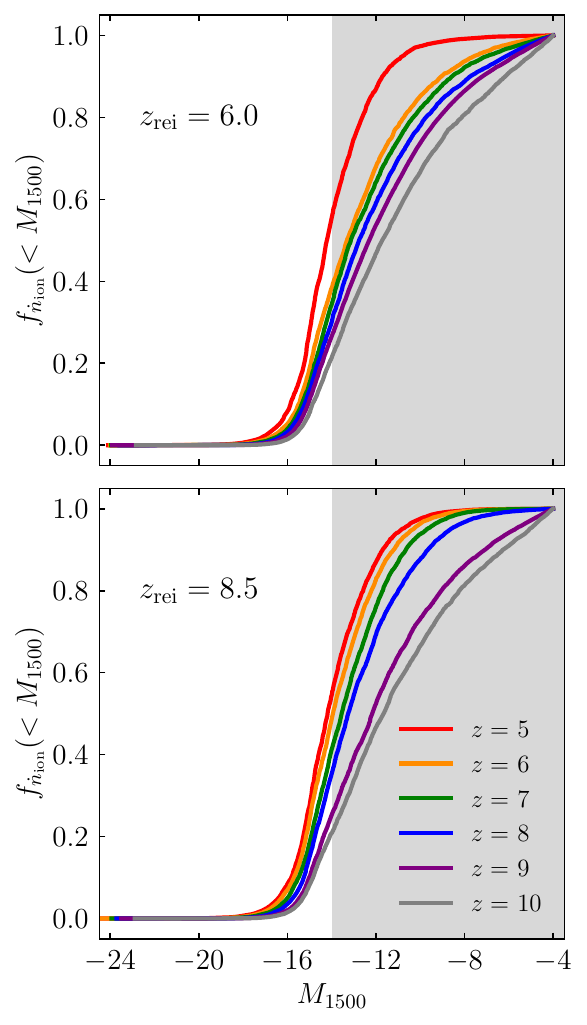}
   }
   \caption{
   The fraction of Lyman continuum flux density contributed by galaxies brighter than a given $\Muv$. {\it Left panels}: the models with $\zrei = 6$ (upper panel) and $8.5$ (lower panel) across $z\in[5,10]$ assuming constant escape fraction of ionizing photons. {\it Right panels}: ionizing photon fraction in the same models, but assuming a strongly luminosity-dependent escape fraction increasing towards fainter galaxies down to $\Muv=-15$ (see Section \ref{sec:frac_ion} for details). The shaded region shows $\Muv>-14$. The figure shows that $\Muv>-14$ galaxies contribute up to $60 \sim 80\%$ at $z>7$ with even ultra-faint galaxies contributing $\approx 10-20\%$.
   }
   \label{fig:fractional_nion}
\end{figure*}

Results presented above show that dwarf galaxies with luminosities beyond the reach of current observations contribute significantly to the total UV emission of galaxies. These galaxies can thus contribute substantially to the reionization of hydrogen in the Universe. However, far UV luminosity at $\lambda=1500\, \aa$ is only a rough proxy of the ionizing radiation produced by galaxies, and the ratio between ionizing luminosity and $L_{1500}$ can vary significantly as a function of the IMF, star formation history, metallicity, and binary fraction \citep[e.g.,][]{Stanway.etal.2016}. 

The conversion from $L_{1500}$ to the ionizing photon emission rate is usually parameterized using $\xiion=\dot{N}_{\rm ion}/L_{1500}$ where $L_{1500}$ is the luminosity per unit frequency in units of $\rm ergs\, s^{-1}Hz^{-1}$, and $\dNion$ is the number of hydrogen-ionizing photons emitted per second. Although $\xiion$ is often considered a constant, it is expected to depend on various factors and can thus vary from galaxy to galaxy as well as with time. Here, instead of adopting a given value of $\xi_{\rm ion}$, we compute the emission rate of ionizing photons, $\dNion$, for each model galaxy using its star formation and metal enrichment history, as described in Section \ref{sec:comp_uv_ion}. We find that the ratio $\dot{N}_{\rm ion}/L_{1500}$ of model galaxies is approximately independent of galaxy luminosity, but there is a substantial scatter due to assumed SFR stochasticity.

In addition to computing $\dNion$, we still need to make assumptions about the value of the escape fraction of ionizing photons from galaxies ($\fesc$) and its scaling with galaxy luminosity. Observational measurements are challenging but have been done for some local galaxies, indicating values of $\lesssim 10\%$ \citep[e.g.,][]{Vanzella.etal.2010,Guaita.etal.2016,Rutkowski.etal.2016,Grazian.etal.2016,Sandberg.etal.2015,Vanzella.etal.2010,Vasei.etal.2016,Flury.etal.2022}. At higher redshifts, escape fractions are found to increase for galaxies with bluer spectra and lower mass \citep[e.g.,][]{Chisholm.etal.2022,Saldana_Lopez.etal.2023} and given that galaxy spectra become bluer on average with increasing redshift \citep{Topping.etal.2022,Cullen.etal.2023}, this implies an increase of $\fesc$ with increasing redshift, reaching values of $\approx 5-30\%$ at $z>6$ \citep[e.g.,][]{Lin.etal.2024,Saxena.etal.2024}. 

On the theoretical side, results of numerical simulations vary significantly from finding a clear trend of increasing $\fesc$ with decreasing galaxy luminosity \citep[e.g.,][]{Wise.etal.2014,Kimm.etal.2017,Anderson.etal.2017} to the opposite trend \citep{Sharma.etal.2016} or $\fesc$ peaking at intermediate masses \citep[e.g.,][]{Rosdahl.etal.2022}. This is not surprising, given that escape fraction depends on the specific exact timing of stellar ionizing and mechanical feedback, the structure of the ISM on a wide range of scale, and other factors \citep[e.g.,][]{Gnedin.etal.2008,Kimm.etal.2014} and thus is extremely sensitive to implementations of star formation, feedback, and numerical resolution of simulations. 

Given this uncertainty, we will adopt two models for $\fesc$ that should reasonably bracket the possible trends. In the first model, we assume a constant $\fesc=5\%$ at all redshifts and luminosities. In the second model, we adopt a strong dependence of $\fesc$ on galaxy luminosity for galaxies with $-21<\Muv<-15$: $\fesc^{\rm eff} = \fesc(\Muv = -21) \times 10 ^ {0.62(\Muv + 21)}$, given by the respective evolution of $\fesc$ and $\xiion$ found in simulations \citet{Anderson.etal.2017} and \citet{Simmonds.etal.2024}; $\fesc$ at $\Muv>-15$ is kept constant at the value given by the above expression at $\Muv=-15$. The latter is done both to avoid extrapolating simulation and observational results, and because large $\fesc$ at lower luminosities likely leads to reionization that is too early \citep[see, e.g.,][]{Lin.etal.2024}. This toy model illustrates how different the results would be in the case of a strong increase of $\fesc$ with decreasing luminosity. It is worth noting though, that the luminosity dependence in this model is likely too strong, as it assumes that both $\fesc$ and $\xiion$ increase with decreasing luminosity, while observations indicate that $\fesc$ and $\xiion$ anti-correlate in high-$z$ galaxies \citep{Saxena.etal.2024} such that their product does not depend strongly on luminosity. As such, this is an overly optimistic model on the contribution of dwarf galaxies to the ionizing photon budget.

For a given $\fesc$ model we order galaxies by their $\Muv$ at a given redshift and compute the cumulative $\fesc$-weighted sum of the ionizing photon emission rate contributed by galaxies brighter than a given $\Muv$. 
Figure~\ref{fig:fractional_nion} shows the fraction of the ionized flux density contributed by galaxies brighter than a given $\Muv$ computed in this way for the two models of $\fesc$ shown in the two panels. For a constant $\fesc$, results are similar to those of $\fuv$: the contribution of faint galaxies with $\Muv>-14$ is $\approx 40-60\%$, which decreases to $\approx 20\%$ at $z=6$. Before reionization at $z \gtrsim \zrei$ even ultra-faint galaxies of $\Muv>-10$ contribute $\approx 10-25\%$ of ionizing photons. For the model that assumes increasing $\fesc$ for fainter galaxies, the contribution of $\Muv>-14$ at $z\geq\zrei$ increases to $\approx 60-75\%$ while $\Muv>-10$ galaxies still contribute $\approx 15-30\%$ before reionization.

\section{Discussion}
\label{sec:discussion}

The results presented in the previous section show that dwarf galaxies beyond the range of luminosities probed by HST and JWST can contribute substantially to the reionization of the Universe.  This is consistent with recent observational results that indicate significant ionizing emissivities of high-$z$ galaxies \citep{Simmonds.etal.2024} and JWST observations of individual dwarf galaxies strongly magnified by cluster lensing \citep{Atek.etal.2024}.
The contribution is smaller than would be estimated from extrapolating the UV LF estimated for galaxies with $\Muv\lesssim-16$, because the slope of the UV LF is predicted to slowly decrease with decreasing luminosity until $\Muv\approx -14$. At fainter luminosities, the slope of our model LFs is approximately constant all the way to $\Muv\approx -4$. This means that if the UV LF can be characterized down to $\Muv\approx -14$ with JWST observations, extrapolation of the LF using slope estimated at the faint end of the measured LF should be accurate. 

The significant contribution of dwarf galaxies to the ionizing photon budget implies that reionization can be achieved with escape fractions of ionizing radiation lower by a factor of up to two than is assumed when the contribution of dwarf galaxies is not taken into account \cite[][]{Finkelstein.etal.2019}. 
If the $\fesc$ increases strongly for fainter galaxies, the implied values of escape fraction can be too high to be consistent with existing observational constraints on the ionization history of our Universe \citep{Lin.etal.2024,Munoz.etal.2024} or with observational estimates of $\fesc$ in galaxies \citep[see discussion in][]{Munoz.etal.2024}. We note that the models that violate observational constraints likely assume $\fesc$ dependence on galaxy luminosity and/or redshifts that are too strong. For example, as noted above, observations indicate that $\fesc$ and $\xiion$ anti-correlate in high-$z$ galaxies \citep{Saxena.etal.2024} such that their product does not depend strongly on luminosity. Thus, models that assume that both $\fesc$ and $\xiion$ increase with decreasing galaxy luminosity independently are likely to have unrealistically high ionizing emissivity for dwarf galaxies.

\subsection{Faint-end UV LF slope}
\label{sec:faint_end_slope}

Several theoretical studies considered model predictions for the faint end of the UV luminosity function with rather diverse results. There is a significant difference between model predictions even at $\M1500\lesssim -13$ \citep[see, e.g., Figures 12 and 13 in][]{Bouwens.etal.2022}. In general, many simulations predict significant flattening or even a turnover of the UV luminosity functions at $\M1500\gtrsim -14$ \citep[e.g.,][]{OShea.etal.2015,Gnedin.2016,Ceverino.etal.2017,Kannan.etal.2022}. Among semi-analytic models, some models predict flattening and turnover at $\M1500<-13$ \citep{Hutter.etal.2021} due to effects of reionization heating, while others predict a relatively steep LF down to fainter magnitudes \citep[e.g.,][]{Yue.etal.2016}.

UV LF at faint magnitudes can be affected by stellar feedback-driven outflows and UV heating due to reionization. Our model incorporates a well-motivated outflow model, with which it reproduces the luminosity function of MW dwarf satellites and many properties of local dwarf galaxies, including their metallicity--stellar mass, gas mass--stellar mass, and Tully-Fisher relations. The outflows in our model result in a gradual decrease of UV LF slope with decreasing luminosity for $\Muv<-14$. As noted above, before reionization the LF slope at $\Muv>-14$ stays approximately the same and corresponds to $dn/dL\propto L^\alpha$ where $\alpha\approx -1.7$, which is quite steep. 
The model presented here also takes into account the effects of reionization on the gas accretion onto dwarf-mass halos, and it shows that such heating does flatten the UV LF at $M_{\rm UV}\gtrsim -13$ but only at redshifts smaller than the reionization redshift $z<z_{\rm rei}$. 

Another process that can affect star formation in small-mass halos is the suppression of gas accretion due to relative velocities of baryons and dark matter \citep{Tseliakhovich.Hirara.2010}. \citet{Williams.etal.2024} evaluated the effect of such motions on the UV LF in their simulations, and found that the relative motions lead to a turnover in the $z=12$ UV LF  at $M_{\rm UV}\gtrsim -13$. This would make the contribution of galaxies at fainter luminosities to the ionizing photon budget negligible -- contrary to our conclusions. However, in their model, this luminosity corresponds to the stellar mass of  
$M_\star\approx 10^5\, M_\odot$ and halo mass of $M_{\rm 200}\approx 10^6\, M_\odot$ below the minimal halo mass of $M_{\rm 200c}\approx 2\times 10^6\,M_\odot$ that can accrete gas and form stars \citep{Nebrin.etal.2023}. 

In our model, however, galaxies  of $M_\star= 10^5\, M_\odot$ and $M_{\rm UV}\approx -12$ occupy halos of mass $M_{200}\approx 3-10\times 10^8\, M_\odot$, or more than two order of magnitude larger. This is consistent with the results of several high-resolution simulations of high-$z$ galaxies shown in Figure 13 of \citet{Cote.etal.2018}. Halos of  $M_{200}=10^6\, M_\odot$, on the other hand, do not form stars and thus do not contribute to UV LF at all in our model. We also note that our model reproduces the UV LFs over a broad range of luminosities and redshifts, as well as various properties of dwarf galaxy population at $z=0$ \citep{Kravtsov.Manwadkar.2022, Kravtsov.Wu.2023}, including the luminosity function of the Milky Way satellite galaxies down to the faintest luminosities \citep{Manwadkar.Kravtsov.2022}.
We thus believe it is likely that in our model, galaxies of a given $M_\star$ would form in halos of correct halo mass. Given that relative baryon and dark matter motions only affect halos of $M_{200}\lesssim 10^6\, M_\odot$, in the context of our model these motions would not affect the UV LF at all at the relevant range of luminosities.

\subsection{UV luminosity density evolution}
\label{sec:uv_density}

\begin{figure}
   \centering {
   \includegraphics[width=0.43\textwidth]{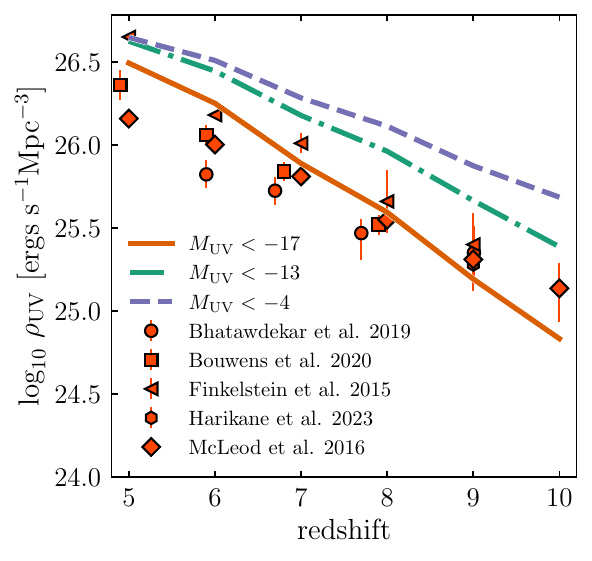}
   }
   \caption{UV ($\lambda=1500\,\aa$) flux density from model galaxies brighter than a given limiting $\Muv$ for $5\leq z\leq 10$. Model predictions for each magnitude limit are shown by the three different lines. Observational estimates from previous studies are shown by points \citet{Bouwens.etal.2023}: circles, \citet{Donnan.etal.2023}: squares, \citet{Oesch.etal.2018}: triangles, \citet{Harikane.etal.2022}: pentagons.
   }
   \label{fig:rho_UV_density}
\end{figure}

The UV flux density $\rhouv$ represents the global probe of how star formation rate density evolves in the Universe, and is the object of many observational and theoretical estimates. We can compute this density using Equation~\ref{eq:f_uv} but with $f_{\rm norm}=1$. Figure~\ref{fig:rho_UV_density} shows our $\rhouv$ values for redshifts $z\in[5, 10]$, integrated from the brightest end to three different UV absolute magnitude limits of $-17$, $-13$, and $-4$.

The three limiting UV magnitudes are significant in different ways: $-17$ is the lowest luminosity of many pre-JWST LF measurements, and is used as a comparison with our model. Figure \ref{fig:rho_UV_density} shows that $\rhouv(z)$ evolution in our model is in reasonable agreement with observational estimates of the UV density measured by integrating UV LF down to $\Muv=-17$  \citep{Harikane.etal.2023, Bouwens.etal.2020, Bhatawdekar.etal.2019, McLeod.etal.2016, Finkelstein.etal.2015}.

The limiting magnitude of $-13$ is an optimistic limit that should be reachable for strongly lensed galaxies observed with JWST. As expected, the corresponding $\rhouv$ values are significantly larger. Lastly, we show $\rhouv$ in our model if we integrate UV LF to $\Muv = -4$ to include UV flux contribution from all dwarf galaxies. The direct contribution of dwarf galaxies to the UV flux density is reflected in the gap between the blue $-4$ and green $-13$ lines. The contribution of dwarf galaxies is significant at higher $z$ and decreases with decreasing redshift. This contribution thus flattens the $\rho_{\rm UV}(z)$ trend produced by dwarf galaxies.

\subsection{Ionizing photon emission rate evolution}
\label{sec:dnion_zevol}

\begin{figure}
   \centering {
   \includegraphics[width=0.49\textwidth]{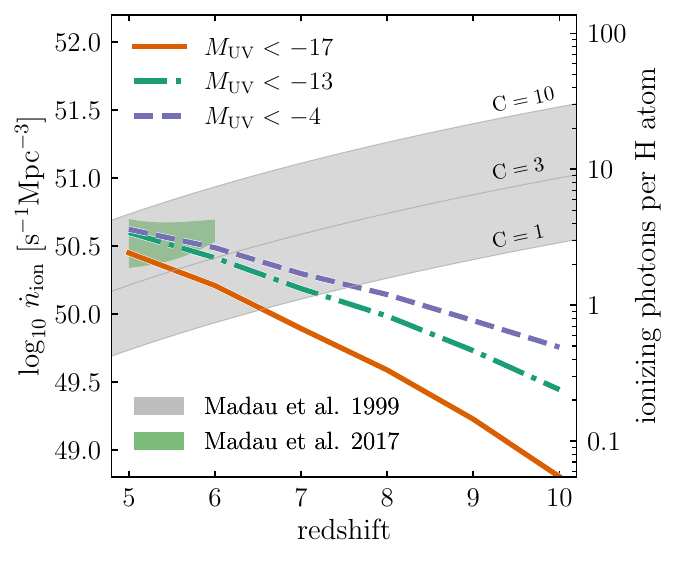}
   }
   \caption{
   The emission rate density of the Lyman continuum photons $\dnion$ for $5\leq z\leq 10$, contributed by galaxies brighter than $-17, -13, -4$ for $\zrei=6$ model assuming a constant escape fraction of $\fesc = 5\%$. The right $y$-axis scale shows a ballpark estimate of the number of ionizing photons per hydrogen atom for each $\dot{n}_{\rm ion}$ value. Also shown are predictions of the estimated $\dnion$ needed to keep the IGM reionized, derived from observations at $z\approx 5-6$ \citep[green band:][]{Fan.etal.2006, Madau.2017} and the theoretical model of \citet{Madau.etal.1999} for clumping factors of $C=1$, $3$, and $10$ (lines, with gray shaded band bracketing models with $C=1$ and $C=10$).
   }
   \label{fig:dnion_zevol}
\end{figure}

The most direct probe of the ionizing photon budget is the ionizing photon emission rate per comoving volume $\dnion$. We obtain this value at each redshift $z\in[5, 10]$ by repeating the calculation in the previous section using Equation~\ref{eq:f_uv} and $f_{\rm norm}=1$, but replacing UV density with the ionizing photon emission rate of each galaxy. Figure~\ref{fig:dnion_zevol} shows the evolution of $\dnion$ with redshift assuming $\fesc=0.05$ for the $\zrei = 6$ model.

As a comparison, we plotted the theoretical estimates of $\dnion$ required to maintain hydrogen reionization at each redshift, given by Equation 26 of \citet{Madau.etal.1999} for three values of the clumping factor $C$ bracketing the range of values measured in cosmological simulations of high-$z$ IGM \citep[e.g.,][]{Gnedin.Ostriker.1997, Madau.etal.1999,Pawlik.etal.2015}. We also added constraints on the ionizing photon emissivity in the IGM, derived from the Gunn–Peterson optical depth measured at $5<z<6$ in the SDSS quasar spectra \citep{Fan.etal.2006, Madau.2017}. 

Finally, the right axis of Figure~\ref{fig:dnion_zevol} shows a ballpark estimate of the number of ionizing photons in the IGM per hydrogen atom corresponding to a given $\dot{n}_{\rm ion}$, calculated as $\dot{n}_{\rm ion}/\bar{n}_{\rm H}\, t_{\rm H,6}$ where $\bar{n}_{\rm H}$ is the mean comoving density of hydrogen atoms and $t_{\rm H,6}=1/H(z=6)$ is the Hubble time at $z=6$ \citep{Mason.etal.2019}. 

Figure~\ref{fig:dnion_zevol} shows that for $\fesc=0.05$ and for typical expected clumping factor $C\approx 3$, hydrogen ionization can be maintained for $z\lesssim 6.5$, which is in qualitative agreement with the current estimates of when reionization occurred \citep[e.g.,][]{Gnedin.Madau.2022}. 
This agreement is approximate and a proper comparison requires model calculations following the ionized hydrogen fraction, which is beyond the scope of this study. 
We note that in addition to the uncertainty of the escape fraction and clumping factor, there are additional uncertainties related to the absorption of ionized photons by the Lyman limit systems \citep[e.g.,][]{Kohler.Gnedin.2007,Furlanetto.Mesinger.2009,McQuinn.etal.2011,Altay.etal.2011,Fan.etal.2024,Georgiev.etal.2024}.

Figure~\ref{fig:dnion_zevol} also shows that the contribution of the galaxies with $\Muv>-13$ will not change the reionization redshift significantly for this case of constant $\fesc$. However, their contribution is larger at higher $z$, such that galaxies of these luminosities can contribute significantly to the formation and growth of high-$z$ ionized bubbles, and increase the total optical depth of ionized gas. This contribution also reduces the required $\fesc$ values for bright galaxies to reionize the Universe at $z\approx 6$.

We also note that if we use the magnitude-dependent $\fesc^{\rm eff}$ expression given in Section~\ref{sec:frac_ion} and Figure~\ref{fig:fractional_nion} (right panels) instead of $\fesc=0.05$, the resulting $\dnion$ from $\Muv<-4$ galaxies is nearly constant at $\log_{10}\dot{n}_{\rm ion}\approx 50.7$, and is within the gray band at all redshifts. It intersects the $C=3$ line at $z\approx 8$. This indicates that for such a strongly mass-dependent escape fraction, the Universe may be reionized too early, in agreement with conclusions of \citet{Munoz.etal.2024}.
On the other hand, the $\dot{n}_{\rm ion}(z)$ evolution in the model of \citet[][see their Fig. 1]{Kulkarni.etal.2019} consistent with various observational constraints and total optical depth of ionized gas has $\log_{10}\dot{n}_{\rm ion}\approx 50.8$ at $z=6$ and $\log_{10}\dot{n}_{\rm ion}\approx 50.3$ at $z=10$. This is only somewhat flatter than the blue dashed line in Figure~\ref{fig:dnion_zevol}, and such evolution can be realized if the escape fraction increases moderately with decreasing galaxy luminosity. 

\section{Summary and conclusions}
\label{sec:summary}

We presented model calculations of the monochromatic ($\lambda=1500\,\aa$) UV luminosity function and Lyman continuum photon flux density function for galaxies at redshifts $z\in[5,10]$ over the entire luminosity range from $\Muv\approx -25$ to $\Muv=-4$. The calculation uses a galaxy formation model shown to reproduce properties of local dwarf galaxies down to the luminosities of the ultra-faint satellites. We focus particularly on the contribution of dwarf galaxies with luminosities $\Muv>-13$ outside the reach of direct observations. 
Our main results and conclusions are as follows:

\begin{itemize}

\item[1.] We characterize the shape of the UV LF predicted by our model over a broad range of absolute magnitudes from $\Muv\approx -25$ to $\Muv=-4$ using a novel method to model the abundance of halos and galaxies of a broad range of mass and luminosity (Section \ref{sec:uv_luminosity_function}). We show that model UV LF can be well described by the modified Schechter form of \citet{Jaacks.etal.2013} at all explored redshifts $z\in [5,10]$. We present the best-fit parameters of this functional form for the model LFs at $z=5, 6, 7, 8, 9, 10$ (Table~\ref{tab:jaacks_params}).

\item[2.] Although the slope of the LFs becomes gradually shallower with decreasing luminosity at $\Muv\lesssim -14$, the UV LF predicted by the model is quite steep at the luminosities beyond the observational limit. After the assumed end of reionization, the UV LF flattens at $M_{1500}\gtrsim -13$ from suppression of gas accretion and star formation in small-mass halos, due to UV heating of intergalactic gas during and after reionization. However, before reionization, the faint end of the LF has slopes at the faintest luminosities as steep as the slope at $M_{1500}\approx -14$. Galaxies fainter than $\Muv=-13$ thus contribute significantly to the UV flux density before reionization at $z>6$ (Figures \ref{fig:fractional_flux} and \ref{fig:rho_UV_density}).

\item[3.] We also compute the ionizing flux of model galaxies brighter than a given absolute magnitude $\Muv$ and show that it can be well described by the same \citet{Jaacks.etal.2013} form. We present the best-fit parameters of this form and an approximation for their evolution with redshift (Appendix \ref{app:nion_params}).

\item[4.] Dwarf galaxies beyond the range of luminosities probed by HST and JWST can contribute substantially to the reionization of the Universe. 
For a constant $\fesc$ the contribution of faint galaxies with $\Muv>-14$ to the ionizing photon budget is $\approx 40-60\%$ at $z>7$, which decreases to $\approx 20\%$ at $z=6$. Before reionization, even ultra-faint galaxies of $\Muv>-10$ contribute $\approx 10-25\%$ of ionizing photons. For the model that assumes a strongly increasing $\fesc$ for fainter galaxies, the contribution of $\Muv>-14$ at $z\geq\zrei$ increases to $\approx 60-75\%$ while $\Muv>-10$ galaxies contribute $\approx 15-30\%$.
\end{itemize}

Our results show that dwarf galaxies play an important role in reionizing the Universe, and may thus significantly aid  the formation and growth of ionizing bubbles at $z>7$. This is in agreement with recent observational estimates of the ionizing flux contributed by dwarf galaxies. Further observational studies using the increasing volume of JWST observations at $z>6$ should improve our understanding of escape fractions and their trends with galaxy properties at high redshifts. 

On the theoretical side, future studies should improve our understanding of the absorption and recombination of photons in the IGM, resulting in a better understanding of the required ionizing photon budget. A natural next step is to improve the treatment of the escape fraction and model the evolution of neutral hydrogen fraction \citep[e.g.,][]{Grieg.etal.2017,Bolan.etal.2022} to constrain and test the model. Observational and theoretical progress in these areas should refine our knowledge of the contribution of dwarf galaxies to the evolution of IGM at $z>5$.

\section*{Acknowledgements}

We are grateful to Nickolay Gnedin, Brant Robertson, Harley Katz and the UChicago structure formation group for useful discussions during this project and to Michael Boylan-Kolchin for catching typo in eq. 4 in earlier versions of this paper. ZW was supported by the University of Chicago CCRF’s Quad Research Scholarship program. AK was supported by the National Science Foundation grants AST-1714658 and AST-1911111 and NASA ATP grant 80NSSC20K0512. 

Analyses presented in this paper were greatly aided by the following free software packages: {\tt NumPy} \citep{NumPy}, {\tt SciPy} \citep{scipy}, {\tt Matplotlib} \citep{matplotlib}, {\tt FSPS} \citep{fsps} and its Python bindings package {\tt Python-FSPS}\footnote{\href{https://github.com/dfm/python-fsps}{\tt https://github.com/dfm/python-fsps}}, {\tt BPASS} stellar population synthesis tables for ionizing luminosity \cite{Byrne.etal.2022}, and {\tt Colossus} cosmology package \citep{colossus}. We have also used the Astrophysics Data Service (\href{http://adsabs.harvard.edu/abstract_service.html}{\tt ADS}) and \href{https://arxiv.org}{\tt arXiv} preprint repository extensively during this project and the writing of the paper.

\section*{Data Availability}

A \GRUMPY model implementation is available at \url{https://github.com/kibokov/GRUMPY}. The data used in the plots within this article are available on request to the corresponding author.


\bibliographystyle{mnras}
\bibliography{reionization}

\appendix

\begin{figure*}
   \centering {
   \includegraphics[width=0.99\textwidth]{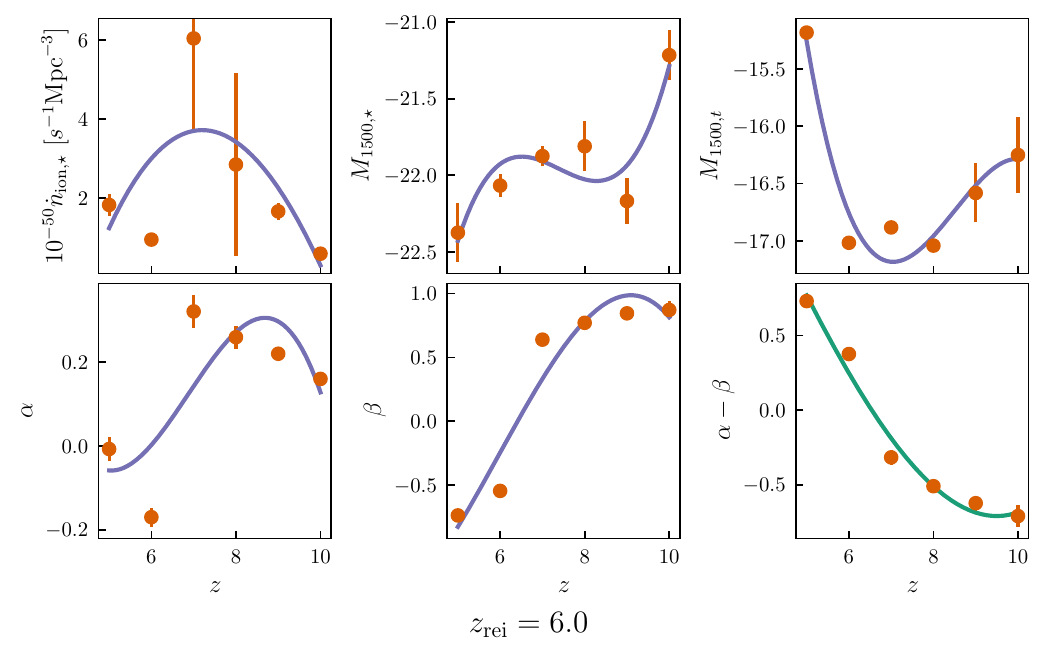}
   \includegraphics[width=0.99\textwidth]{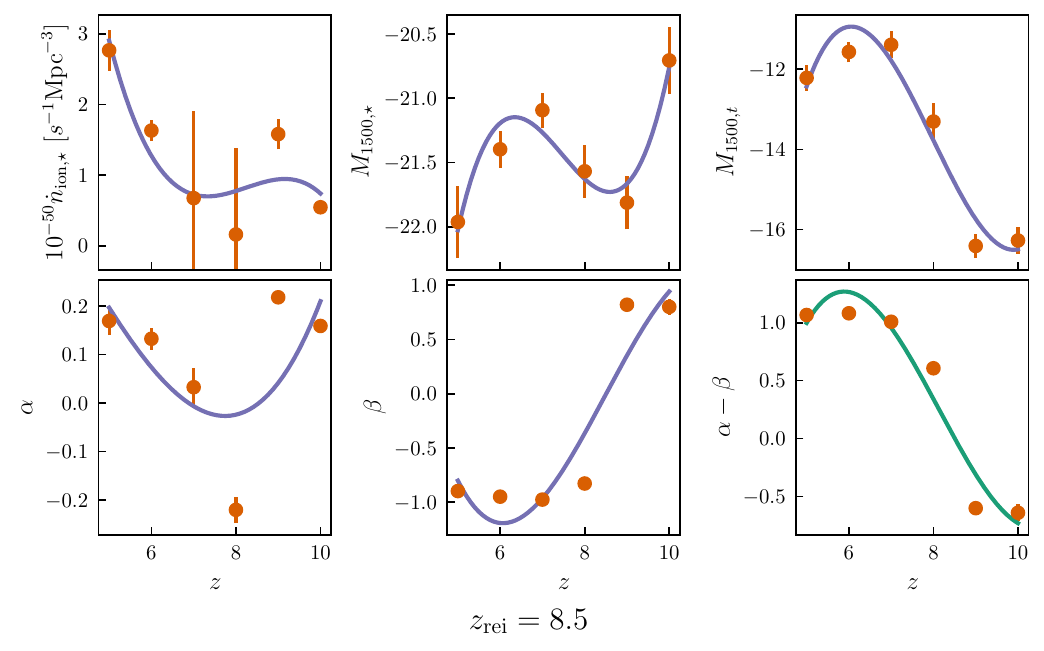}
   }
   \caption{
   Coefficients of the third-order polynomial fit approximation to the evolution of the parameters of the \citet{Jaacks.etal.2013} analytical form to $\dot{n}_{\rm ion}(\Muv)$ at $5\leq z\leq 10$. {\it Top panel\/}: model with $\zrei = 6.0$. {\it Bottom panel}: model with $\zrei = 8.5$. The polynomial coefficients are presented in Table~\ref{tab:nion_poly_fit}.
   }
   \label{fig:nion_polynomial}
\end{figure*}

\section{$\nion$ functional fit parameter}
\label{app:nion_params}

We construct an ionizing flux function as a function of $\Muv$, $\dot{n}_{\rm ion}(\Muv)$ similarly to how we estimate model UV LF (Section \ref{sec:comp_uv_ion}). Namely, we estimate $\dot{n}_{\rm ion}(\Muv)$ as a weighted histogram of halos in a box of a given comoving size $L_{\rm box}$ in bins of $\Muv$ with weights given by $\dot{N}_{\rm ion}/(fL_{\rm box}^3)$, where $\dot{N}_{\rm ion}$ is the Lyman continuum photom emission rate of each model galaxy and $f=f(M_{\rm 200c})$ is the fraction of selected halos in a given box. 

This function can also be approximated analytically by the same modified Schechter function from \citet{Jaacks.etal.2013}, in which  $\phis$ in Equation~\ref{eq:jaacks} is replaced with ${\dot n}_{\rm ion, \star}$. Fit parameters for the ionizing photon flux are shown in Table~\ref{tab:nion_params}, analogous to $\M1500$ UV LF parameters presented earlier in Table~\ref{tab:jaacks_params}.

\begin{table}
    \centering
	\caption{
 Best-fit parameters for \citet{Jaacks.etal.2013}'s modified Schechter function to the ionizing flux function $\dot{n}_{\rm ion}$ at $5\leq z\leq 10$.
 }
	\begin{tabular}{cccccc}
		\hline\hline\\[-2mm]
		$z$ & $\log_{10} 10^{-50}{\dot n}_{\rm ion, \star}$ & $M_{1500, \star}$ & $M_{1500, t}$ & $\alpha$ & $\beta$ \\[1mm]
		\hline\\[-2mm]
        \multicolumn{6}{c}{$\zrei = 6.0$} \\[1mm]
            5 & 0.262 & -22.38 & -15.18 & -0.007 & -0.738  \\
            6 & -0.021 & -22.07 & -17.02 & -0.170 & -0.546  \\
            7 & 0.781 & -21.88 & -16.88 & 0.322 & 0.639  \\
            8 & 0.455 & -21.81 & -17.04 & 0.260 & 0.770  \\
            9 & 0.222 & -22.17 & -16.58 & 0.220 & 0.844  \\
            10 & -0.225 & -21.22 & -16.25 & 0.160 & 0.871 \\[1mm]
        \multicolumn{6}{c}{$\zrei = 8.5$} \\[1mm]
            5 & 0.442 & -21.96 & -12.22 & 0.170 & -0.897  \\
            6 & 0.212 & -21.40 & -11.57 & 0.133 & -0.949  \\
            7 & -0.171 & -21.09 & -11.40 & 0.033 & -0.976  \\
            8 & -0.794 & -21.57 & -13.31 & -0.221 & -0.828  \\
            9 & 0.199 & -21.81 & -16.41 & 0.218 & 0.820  \\
            10 & -0.263 & -20.71 & -16.27 & 0.159 & 0.800  \\[1mm]
    \hline
    \label{tab:nion_params}
    \end{tabular}
\end{table}

Note that this functional form is not used in any of the analyses in this paper; we use the actual estimate of the function from the model.
The approximation is provided here so that one can compute LyC photon budget or model the ionization history of the Universe by evolving $\dot{n}_{\rm ion}$ over time. To this end, we also provide an approximation for how best-fit parameters of the functional form evolve with redshift for $5\leq z\leq 10$. Namely, we approximate the evolution of each parameter using third-order polynomials ${\rm param} = a_0 + a_1 z + a_2 z^2 + a_3 z^3$. Figure~\ref{fig:nion_polynomial} shows the polynomial fits for each parameter, while Table~\ref{tab:nion_poly_fit} presents the best-fit values of $a_i$ coefficients.\\[3mm]

\begin{table}
    \centering
	\caption{
 Coefficients of the third-order polynomial fit approximation to the evolution of the parameters of the \citet{Jaacks.etal.2013} approximation to $\dot{n}_{\rm ion}(\Muv)$ with redshift at $5\leq z\leq 10$.
 }
	\begin{tabular}{ccccc}
		\hline\hline\\[-2mm]
        \multicolumn{5}{c}{$\zrei = 6.0$} \\[1mm]
         Parameters & $a_3$ & $a_2$ & $a_1$ & $a_0$ \\[1mm]
         $10^{-50}{\dot n}_{\rm ion, \star}$ & 0.015 & -0.801 & 9.216 & -26.682 \\
         $M_{1500, \star}$ & 0.058 & -1.284 & 9.356 & -44.352 \\
         $M_{1500, t}$ & -0.072 & 1.830 & -15.108 & 23.492 \\
         $\alpha$ & -0.015 & 0.315 & -2.013 & 4.045 \\
         $\beta$ & -0.021 & 0.367 & -1.594 & 0.520 \\[1mm]
        \multicolumn{5}{c}{$\zrei = 8.5$} \\[1mm]
         Parameters & $a_3$ & $a_2$ & $a_1$ & $a_0$ \\[1mm]
         $10^{-50}{\dot n}_{\rm ion, \star}$ & -0.080 & 1.988 & -16.183 & 44.162 \\
         $M_{1500, \star}$ & 0.103 & -2.303 & 16.817 & -61.377 \\
         $M_{1500, t}$ & 0.196 & -4.701 & 35.313 & -96.041 \\
         $\alpha$ & 0.003 & -0.039 & 0.001 & 0.750 \\
         $\beta$ & -0.042 & 1.057 & -8.236 & 19.151 \\[1mm]
    \hline
    \label{tab:nion_poly_fit}
    \end{tabular}
\end{table}


\end{document}